\documentclass[letterpaper, 10 pt, conference]{ieeeconf}
\IEEEoverridecommandlockouts 
\pdfoutput=1 

\usepackage{times} 
\usepackage{amsfonts}
\usepackage{stmaryrd}
\usepackage{bm}
\usepackage{booktabs}
\usepackage{mathtools} 

\usepackage{subfig}

\usepackage{etoolbox}
\makeatletter
\patchcmd{\@makecaption}
  {\\}
  {.\ }
  {}
  {}
\makeatletter
\patchcmd{\@makecaption}
  {\scshape}
  {}
  {}
  {}
\makeatother

\title{Data-Driven Discrepancy Modeling in Higher-Dimensional State Space via Coprime Factorization}

\author{Sourav Sinha and Mazen Farhood
\thanks{The authors are with the Kevin T. Crofton Department of Aerospace and Ocean Engineering, Virginia Tech, Blacksburg, VA 24061, USA. Email: \{srvsinha, farhood\}@vt.edu.}\thanks{This work was supported by the Army Research Office under Contract No. W911NF-21-1-0250.}}

\begin{document}

\maketitle
\thispagestyle{empty}
\pagestyle{empty}

\begin{abstract}
This work provides a data-driven framework that combines coprime factorization with a lifting linearization technique to model the discrepancy between a nonlinear system and its nominal linear approximation using a linear time-invariant (LTI) state-space model in a higher-dimensional state space. In the proposed framework, the nonlinear system is represented in terms of the left coprime factors of the nominal linear system, along with perturbations  modeled as stable, norm-bounded LTI systems in a higher-dimensional state space using a deep learning approach. Our method builds on a recently proposed parametrization for norm-bounded systems, enabling the simultaneous minimization of the ${H}_\infty$-norm of the learned perturbations. We also provide a coprime factorization-based approach as an alternative to direct methods for learning lifted LTI approximations of nonlinear systems. In this approach, the LTI approximations are obtained by learning their left coprime factors, which remain stable even when the original system is unstable. The effectiveness of the proposed discrepancy modeling approach is demonstrated through multiple examples.
\end{abstract}

\section{Introduction}  \label{sec_intro}

Most physical processes exhibit nonlinear dynamics. However, in many applications, it is often desirable to construct a linear time-invariant (LTI) system that approximates the behavior of the nonlinear dynamical system over the operating envelope of interest. Such a representation reduces the model complexity and enables the application of well-established methodologies developed for linear systems. For instance, control techniques developed for linear systems \cite{gahinet1994linear,scherer1997multiobjective} can be leveraged to design controllers for nonlinear systems with mathematical stability and performance guarantees over the considered operating envelopes. In many scenarios, first-principles-based mathematical abstractions of physical processes are available. Traditional linearization techniques can then be used to generate an approximate linear model. However, such models are typically valid only locally, and, as a result, may not fully capture the nonlinear system’s behavior across the entire envelope of interest. Additionally, the abstraction itself may not accurately represent the actual physical process. Therefore, it is important to account for the discrepancy between the predictions of the nominal linear model and observed measurements. 

The data-driven approximation of discrepancy models has been addressed previously in works such as \cite{sinha2021lft, kaheman2019learning}. In these works, the discrepancy is modeled by introducing an error term in the original state equation. The error term is given in terms of polynomial or, more generally, nonlinear scalar-valued candidate functions that are chosen a priori. Consequently, this approach yields at best a linear parameter-varying approximation of the original nonlinear dynamics. Additionally, these approaches frame the learning task as a regression problem, where the focus is on minimizing the prediction error over just one time step. When the system is unstable, the responses of the nonlinear system and the nominal linear model can quickly diverge for the same input. This leads to an exponential amplification of the errors, thereby posing significant challenges when learning models aimed at minimizing long-term prediction errors.

In this work, discrepancy models are approximated as LTI systems. The discrepancy between the true and nominal models arises due to the neglected nonlinearities and the imperfections in the mathematical abstraction; thus, modeling this discrepancy using an LTI system can be challenging. To address this difficulty, we leverage a lifting technique as in \cite{williams2015data, korda2018linear, narasingam2019koopman, otto2019linearly} to generate an LTI state-space system in a higher-dimensional space that can reasonably approximate the nonlinear discrepancy model within the envelope of interest. We propose a coprime factorization-based framework to learn the LTI discrepancy model. In this framework, the linear approximation of the nonlinear dynamics is represented as an interconnection of the left coprime factors of the known nominal system and (stable) norm-bounded coprime factor perturbations, which are modeled as LTI systems in a higher-dimensional state space. A deep learning approach is used to simultaneously learn the ``lifting" function, which maps the system state to a ``lifted'' state in a higher-dimensional space, and the system matrices of the coprime factor perturbations in the lifted state space. 

The learning process involves minimizing a loss function that measures the long-term prediction error over any specified number of time steps.
The coprime factorization-based approach provides the advantage of improved numerical stability for long-term predictions during training, as the coprime factor perturbations are inherently stable. Furthermore, we incorporate a regularization term in the loss function to minimize the ${H}_\infty$-norm of the learned perturbations. This is done to ensure that the perturbed system stays close to the first-principles-based model in a gap metric sense \cite{skogestad2005multivariable, georgiou1989optimal}. Various works have addressed the norm-regularized or norm-constrained model estimation problem; see, for example, \cite{verhoek2023learning, hara2020learning, hara2021gain, dahdah2022system}. In this work, we build  on the direct parametrization approach proposed in \cite{verhoek2023learning}, which enables us to solve an unconstrained learning problem. Finally, we apply the proposed approach to model the discrepancies between the nonlinear dynamics of a simple pendulum, a Van der Pol oscillator, and a three-degrees-of-freedom (3-DOF) unmanned aircraft system (UAS) and their respective~linearized~models.

In works such as \cite{korda2018linear,proctor2018generalizing,narasingam2019koopman,folkestad2020extended,mamakoukas2019local}, purely data-driven approaches are used to directly learn LTI approximations of the nonlinear dynamics in a higher-dimensional state space. However, the discrepancy modeling approach is appealing because it leverages information already available from the first-principles-based model. Moreover, purely data-driven approaches typically require a substantial amount of data for effective training and are prone to overfitting. Consequently, such models may fail to generalize beyond the specific envelope or range of conditions within which they were trained. In contrast, first-principles-based models tend to generalize better to unseen data and scenarios, making them more reliable across a broader operating envelope. Through the 3-DOF UAS example, we demonstrate the benefits of the discrepancy modeling approach over purely data-driven modeling methods, especially when limited data is available for training the model.

The rest of the paper is organized as follows: Section~{2} presents the notation, outlines the problem statement, and provides some needed background on the lifting approach and coprime factorization. In Section~{3}, we delve into the data-driven estimation of the discrepancy model via coprime factorization. The proposed approach is applied to three numerical examples in Section~{4}. Section~{5} offers concluding remarks and directions for future work. Finally, an appendix is included, which provides an additional coprime factorization-based result for directly learning lifted LTI approximations of nonlinear systems.

\section{Preliminaries}  

\subsection{Notation}

The sets of non-negative integers, real vectors of dimension $n$, and real $n\times m$ matrices are denoted by $\mathbb{Z}_+$, $\mathbb{R}^n$, and $\mathbb{R}^{n\times m}$, respectively. The identity matrix of size $n\times n$ and the zero matrix of size $n\times m$ are denoted by $I_n$ and $0_{n\times m}$, respectively. Subscripts are dropped when the size is clear from the context. For $v \in \mathbb{R}^n$, $\text{diag}(v)$ denotes an $n\times n$ diagonal matrix where the elements of $v$ are placed along the main diagonal. The trace, $i^{th}$ eigenvalue, and maximum eigenvalue of a square matrix $X \in \mathbb{R}^{n \times n}$ are denoted by $\text{Tr}(X)$, $\lambda_i(X)$, and $\lambda_{\max}(X)$, respectively. 
For any nonnegative integers $m$, $n$, with $n > m$, the sequence $(d(m),\,d(m+1),\,\dots,\,d(n))$ is denoted as $\{d(j)\}_{j=m}^n$.

\subsection{Problem Statement}

We consider a nonlinear dynamical system described by the following discrete-time state equation:
\begin{equation}
    x(k+1) = f\left(x(k),u(k)\right), \quad x(0) = x_0,
    \label{eqn_NLdyn}
\end{equation}
where $x \in \mathbb{X} $ and $u \in \mathbb{U} $ are the state and input of the system, respectively,  with $x(k)\in\mathbb{R}^n$ and $u(k)\in\mathbb{R}^m$ denoting their values at discrete-time instant $k$. The function $f : \mathbb{R}^n \times \mathbb{R}^m \rightarrow \mathbb{R}^n$ defines the nonlinear dynamics of the system. The initial value $x_0$ of the state is considered uncertain. The sets $\mathbb{X}$ and $\mathbb{U}$ together define the operating envelope of interest.  Without loss of generality, we assume that the nonlinear system has an equilibrium at the origin. A non-zero equilibrium can be shifted to the origin through a change of variables.

Suppose that a nominal LTI system $P$ approximating the true nonlinear dynamics around an equilibrium is available and is defined as follows:
\begin{equation}
    {x}(k+1) \approx A_P {x}(k) + B_Pu(k),  \quad x(0) = x_0.  \\
   \label{eqn_NOMsys}
\end{equation}
This model may be obtained by linearizing a mathematical abstraction of the nonlinear system about the equilibrium. Our objective is to model the discrepancy between this nominal LTI system and the true nonlinear dynamics using a dynamic LTI perturbation such that the perturbed linear system, denoted by $G$, approximates the nonlinear system behavior with improved accuracy. We assume that multiple time histories of the nonlinear system's state and input are available for learning the discrepancy model. This data can be generated through experiments or, if the nonlinear system is known, by simulating the nonlinear system with non-zero initial conditions and carefully designed inputs to ensure that the operating envelope of interest $(\mathbb{X},\, \mathbb{U})$ is not violated.

\subsection{Linear Lifted Representations of Nonlinear Systems}

Koopman operator-based methods have emerged as a powerful paradigm for data-driven linear lifted approximations of nonlinear dynamical systems. The Koopman operator \cite{koopman1931hamiltonian} governs the temporal evolution of observables, which are functions of the system’s state. Although the underlying dynamics are nonlinear, the Koopman operator is linear in the space of observables; however, it is typically infinite dimensional. For prediction and control applications, data-driven techniques like extended dynamic mode decomposition \cite{williams2015data} and deep learning \cite{lusch2018deep,otto2019linearly,kaiser2021data}, which are tied to Koopman operator theory, are commonly employed to obtain finite-dimensional linear approximations of nonlinear systems. The key idea involves learning a nonlinear mapping that ``lifts” the system state to a higher-dimensional space of observables, where the system dynamics become approximately~linear.

As in \cite{korda2018linear,narasingam2019koopman,proctor2018generalizing,folkestad2020extended,mamakoukas2019local}, the higher-dimensional LTI representation of the nonlinear system in \eqref{eqn_NLdyn} can be~defined~as 
\begin{equation}
\!\!\! G \colon \!\! \begin{cases}
    {z}(k+1) = A_G {z}(k) + B_Gu(k),  \, z(0) = \Phi(x_0),  \\
     \hspace{5.8mm} {x}(k) = C_G {z}(k),  
  \end{cases}
  \label{eqn_LTIsys}
\end{equation}
where $\Phi : \mathbb{R}^n \rightarrow \mathbb{R}^N$ $(N \gg n)$ is the lifting function and $z(k) = \Phi(x(k)) \in \mathbb{R}^N$ represents the ``lifted" state vector at discrete-time instant $k \in \mathbb{Z}_+$.

\begin{figure}[t]
\begin{center}
\centerline{\includegraphics[width=0.46\textwidth]{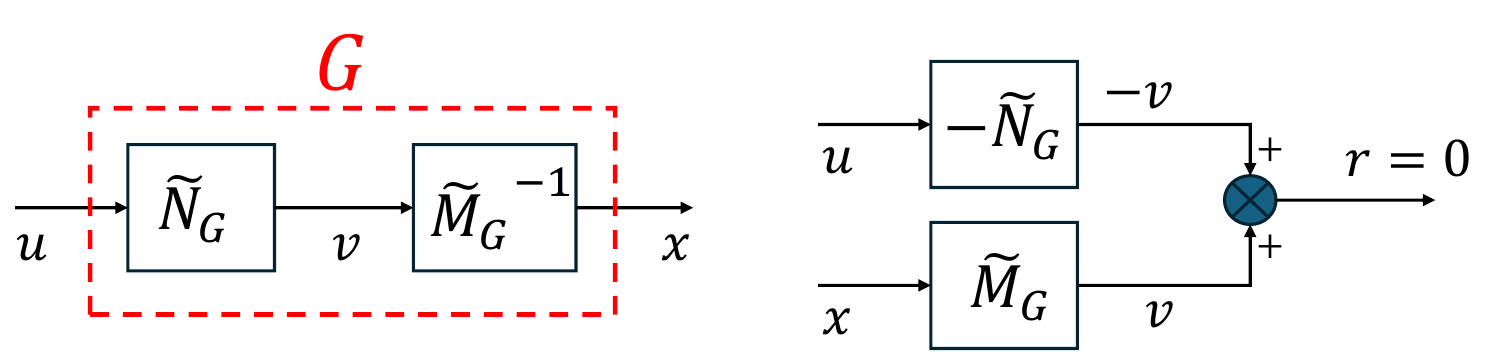}}
\caption{Left coprime factor representation.}
\label{fig_LCF}
\end{center}
\end{figure}

\subsection{Coprime Factorization} \label{sec_CFR}

The (potentially unstable) system $G$, defined in \eqref{eqn_LTIsys}, can be factored as ${G} = {\tilde{M}}_G^{-1}\tilde{N}_G$, where $\tilde{M}_G$ and $\tilde{N}_G$ are the left coprime factors of $G$. These factors are stable even when the system $G$ itself may be unstable. Suppose $(C_G,\,A_G)$ is detectable and $L_G$ is a matrix such that $A_G+L_GC_G$ is stable, i.e., all its eigenvalues lie inside the unit circle. Then, the minimal state-space realization of a left coprime factorization of $G$ is given by \cite{zhou1996robust}
\begin{equation}
   \left[-\tilde{N}_G ~~ \tilde{M}_G\right] \coloneqq \left [ \begin{array}{c|cc}  A_G + L_GC_G & -B_G & L_G \\ \midrule C_G & 0 & I \end{array} \right].
    \label{eqn_lcf}
\end{equation}
There exist multiple solutions to the gain matrix $L_G$ that can be generated using the pole placement method or by solving discrete-time algebraic Riccati equations.
Let $x$ be the output of $G$ for any arbitrary initial state $z(0) \in \mathbb{R}^N$ and input $u \in \mathbb{U}$, then the left coprime factorization satisfies the following (see Figure~\ref{fig_LCF}):
\begin{equation}
    \begin{bmatrix} \tilde{z}(k+1) \\ r(k) \end{bmatrix} = 
    \begin{bmatrix}  A_G + L_GC_G & -B_G & L_G \\  C_G & 0 & I \end{bmatrix} \begin{bmatrix} \tilde{z}(k) \\ u(k) \\ x(k) \end{bmatrix}, 
    \label{eqn_lcfss}
\end{equation}
where $r(k) = 0$ for all $k \in \mathbb{Z}_+$ and the vector $\tilde{z}(k) \in \mathbb{R}^N$ represents the state of the coprime factorization at discrete instant $k$. Inspecting the output equations of \eqref{eqn_LTIsys} and \eqref{eqn_lcfss}, it is not difficult to see that $\tilde{z}(k) = -z(k)$ for all $k\in \mathbb{Z}_+$. In Appendix~\ref{sec_app}, we present a method that can be used to directly learn the left coprime factorization of $G$ using data. The linear lifted approximation can then be constructed from the learned coprime factors.

\section{Data-Driven Approximations of Discrepancy Models}

\begin{figure}[t]
\begin{center}
\centerline{\includegraphics[width=0.46\textwidth]{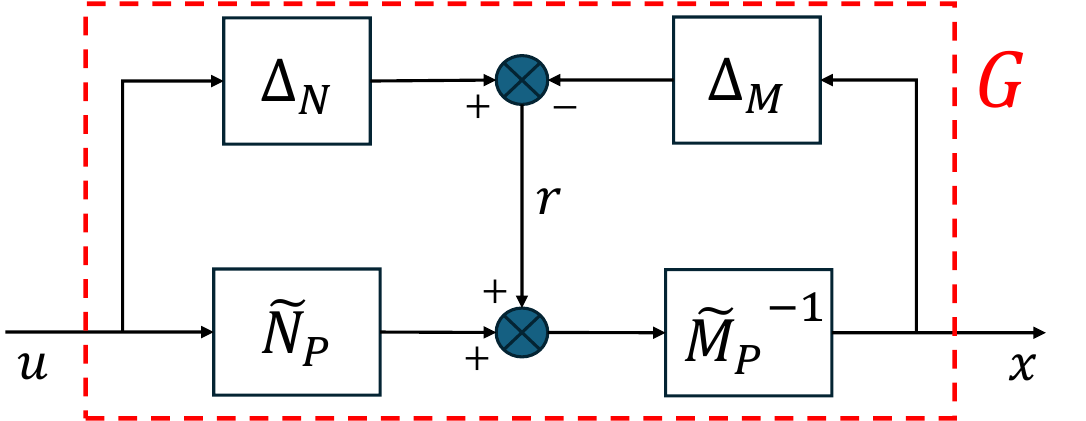}}
\caption{Left coprime factor perturbed system.}
\label{fig_LCFpert}
\end{center}
\end{figure}

Given the nominal LTI approximation $P$ of the nonlinear system, we model the improved, higher-dimensional, linear approximation $G$ as a 
left coprime factor perturbed system 
\begin{equation}
    G =  {\tilde{M}}_G^{-1}\tilde{N}_G = (\tilde{M}_P+\Delta_M)^{-1}(\tilde{N}_P+\Delta_N), \label{eqn_CFpert}
\end{equation}
where $\tilde{M}_P$ and $\tilde{N}_P$ are the left coprime factors of $P$, i.e., $P = \tilde{M}_P^{-1}\tilde{N}_P$, and $\Delta_M$, $\Delta_N$ are stable, norm-bounded, dynamic perturbations that account for the discrepancy between the responses of the true nonlinear system and the nominal linear system \cite{zhou1996robust}. As the nominal system is observable with $C_P = I$, the state-space representation of its coprime factorization $[-\tilde{N}_P ~~ \tilde{M}_P]$ can be expressed as 
\begin{equation}
    \begin{bmatrix} \tilde{x}(k+1) \\ r(k) \end{bmatrix} = 
    \begin{bmatrix}  A_P + L_P & -B_P & L_P \\  I & 0 & I \end{bmatrix} \begin{bmatrix} \tilde{x}(k) \\ u(k) \\ x(k) \end{bmatrix},
    \label{eqn_lcfP}
\end{equation}
where $\tilde{x}(0) = -x_0$. Here, $u$ and $x$ are the input and output of the true nonlinear system, respectively, and $r$ represents the residual error, which is non-zero due to the discrepancy between the true nonlinear system and the nominal system $P$. The perturbed linear system $G$ will be an exact representation of the nonlinear system if the output of the system $[\Delta_N ~~ -\Delta_M]$, for the same input pair $(u,\,x)$, matches the residual $r$ generated by the nominal system, as shown in Figure~\ref{fig_LCFpert}.

\subsection{Modeling the Coprime Factor Perturbations}

We describe the (unknown) dynamic coprime factor perturbations in a higher-dimensional space by the following state-space equations:
\begin{equation}
 [\Delta_N ~~ -\Delta_M]  \colon \begin{cases}
     \tilde{z}_\Delta(k+1) = A_\theta \tilde{z}_\Delta(k) + B_\theta \tilde{u}(k),\\
     \hspace{9.25mm} \hat{r}(k) = C_\theta \tilde{z}_\Delta(k),
  \end{cases}
  \label{eqn_CFpertsys}
\end{equation}
where $\tilde{z}_\Delta(0) = \Psi(x_0)$ with $\Psi : \mathbb{R}^n \rightarrow \mathbb{R}^N$ as the lifting function. Here, $\tilde{u}(k) = [u(k)^T,\,x(k)^T]^T$, $\hat{r}$ is the approximation of the residual $r$, and $\tilde{z}_\Delta(k) \in \mathbb{R}^N$ for $k \in \mathbb{Z}_+ \!\setminus\! \{0\}$ approximates the lifted state vector $\Psi(x(k))$. 
From \eqref{eqn_lcfP} and \eqref{eqn_CFpertsys}, we get the following state-space realization for the left coprime factorization of~$G$:
\begin{equation}
 \!\!\!  \left[-\tilde{N}_G ~~ \tilde{M}_G\right] \coloneqq \left [ \begin{array}{cc|cc}  A_P + L_P & 0 & -B_P & L_P \\ 0 & A_\theta & B_{\theta_1} & B_{\theta_2}\\  \midrule I & -C_\theta & 0 & I \end{array} \right]\!\!,
    \label{eqn_lcfG}
\end{equation}
where $B_{\theta_1}$ and $B_{\theta_2}$ are obtained by partitioning $B_{\theta}$ such that $B_{\theta}\tilde{u}(k) = B_{\theta_1} u(k) + B_{\theta_2} x(k)$. Comparing \eqref{eqn_lcf} and \eqref{eqn_lcfG}, the state-space matrices of the system $G$ are recovered as follows:
\begin{equation}
\begin{split}
    &A_G = \begin{bmatrix} A_P & L_P C_\theta \\ -B_{\theta_2} & A_\theta + B_{\theta_2}C_\theta  \end{bmatrix},\\
    &B_G = \begin{bmatrix} B_P \\ -B_{\theta_1}   \end{bmatrix}, \quad C_G = \begin{bmatrix} I & -C_{\theta}   \end{bmatrix}.
\end{split}
\label{eqn_sstransformation}
\end{equation} 
The initial state of $G$ is then given by 
$$z(0) = -\tilde{z}(0) = -[-x_0^T, ~ \tilde{z}_\Delta(0)^T]^T =  [x_0^T, ~ -\Psi(x_0)^T]^T.$$

\begin{figure}[t]
\begin{center}
\centerline{\includegraphics[width=0.46\textwidth]{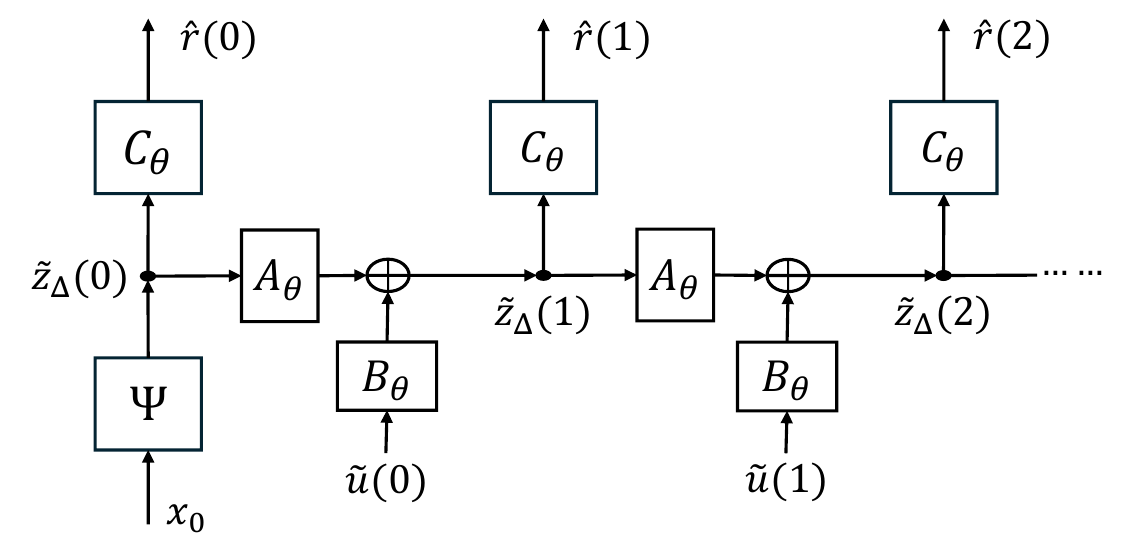}}
\caption{RNN architecture for learning.}
\label{fig_RNN}
\end{center}
\end{figure}

\subsection{Learning Approach} \label{sec_learning}

Consider a training dataset $\mathcal{D}$ formed by multiple finite-horizon trajectories of \eqref{eqn_NLdyn} in the considered operating envelope, each denoted as $\{ \left(x(j),\, u(j) \right) \}_{j=0}^T$. For each trajectory in the dataset, the residual signal $\{  r(j) \}_{j=0}^T$ is generated by simulating the left coprime factorization of $P$, defined in \eqref{eqn_lcfP}, with $(u,\,x)$ as input and $-x_0$ as initial state.
The lifting function $\Psi$ is modeled using a deep neural network parameterized by $\eta$ (network weights). We remove the bias term from each layer to ensure that $\Psi(0) = 0$, which implies that the learned system also has an equilibrium at the origin. The chosen activation function $\sigma$ must also satisfy $\sigma(0) = 0$ for this condition to hold. Commonly used activation functions meeting this criterion include tanh, ReLU, ELU, etc. The architecture of the lifting function, which is determined by the number of hidden layers, the width of the hidden layers, and the choice of activation function, is treated as a learning hyperparameter.

The function $\Psi$ and matrices $(A_\theta,\, B_\theta,\, C_\theta)$ can be learned simultaneously by minimizing 
\begin{equation}
    \mathcal{L}_\mathcal{D}(\eta,\,\theta) = \mathcal{L}_{\text{pred}} + \beta \mathcal{L}_{\text{dyn}}, \label{eqn_lossfn}
\end{equation}
where the positive weight $\beta$ is a learning hyperparameter and $(\eta,\,\theta)$ are the unknown variables that parameterize the lifting function and the state-space matrices. The losses $ \mathcal{L}_{\text{pred}}$ and $ \mathcal{L}_{\text{dyn}}$ denote the mean \textit{prediction} and \textit{linear dynamics} losses, respectively, computed over all the trajectories in the training dataset. For a single trajectory in $\mathcal{D}$, these losses are defined as follows:
\begin{align*}
    \mathcal{L}_{\text{pred}} &=  \frac{1}{T+1}\sum_{k=0}^{T}\lVert \hat{r}(k) - {r}(k)\lVert_{\text{MSE}}, \\
    \mathcal{L}_{\text{dyn}} &= \frac{1}{T}\sum_{k=1}^{T}\lVert \tilde{z}_\Delta(k) - \Psi(x(k))\lVert_{\text{MSE}},
    \label{eqn_loss} 
\end{align*}
where MSE refers to mean squared error. Here, $\tilde{z}_\Delta$ and $\hat{r}$ are computed by recursively applying \eqref{eqn_CFpertsys} akin to a recurrent neural network (RNN); see Figure~\ref{fig_RNN}. 
During learning, the stability of the system can be explicitly enforced using direct parametrizations of the state matrix, as proposed in works such as \cite{verhoek2023learning, fan2022learning, di2023stable}. In this work, however, the stability requirement is implicitly satisfied by learning norm-bounded perturbations, specifically through minimizing the $\mathcal{H}_\infty$ norm of $\left[\Delta_N ~~ -\Delta_M\right]$ during the learning process.

\subsection{$\mathcal{H}_\infty$-norm Regularization} \label{sec_hinf}

In \cite{verhoek2023learning}, a direct parametrization of the state-space matrices is proposed for $\mathcal{H}_\infty$-norm-bounded linear systems. We adopt this parametrization to learn norm-bounded coprime perturbations using unconstrained learning techniques.
Consider a stable, discrete-time, LTI system $G$, with a realization $(A,B,C,D)$, where $A \in \mathbb{R}^{n_x \times n_x}$, $B \in \mathbb{R}^{n_x \times n_u}$ and  $C \in \mathbb{R}^{n_y \times n_x}$, $D \in \mathbb{R}^{n_y \times n_u}$ are the state-space matrices defining the system's state and output equations, respectively. 

\textit{Direct parametrization \cite{verhoek2023learning}:}
For any $V \in \mathbb{R}^{n_x \times n_x}$, $d=[d_1,\ldots,d_{n_x}]^T \in \mathbb{R}^{n_x}$, $X \in \mathbb{R}^{\bar{n} \times \bar{n}}$, $Y \in \mathbb{R}^{\bar{n} \times \bar{n}}$, $Z \in \mathbb{R}^{\underline{n} \times \bar{n}}$, and $\gamma > 0$, where $\bar{n} = n_x + \text{min}(n_y,\,n_u)$ and $\underline{n} = \lvert n_y - n_u \rvert$, the LTI system $G$, with state-space matrices parametrized as
\begin{equation}
    \left[ \begin{array}{cc}  A & B \\  C & D \end{array} \right] =
    \left[ \begin{array}{cc}  Q\Lambda^{-1} & 0 \\ 0 & I \end{array} \right] {M} \left[ \begin{array}{cc}  \Lambda Q^T & 0 \\ 0 & \gamma I \end{array} \right],
    \label{eqn_ssparametrization}
\end{equation}
is stable and satisfies $\lVert G \rVert_\infty \leq \gamma$. Here, $Q = (I-V+V^T)(I+V-V^T)^{-1}$, $\Lambda = \text{diag}([e^{d_1},\ldots, e^{d_{n_x}}]^T)$,  and $M = \bar{M}$ if $n_y \geq n_u$; otherwise, $M = \bar{M}^T$, where, for some constant~$\epsilon > 0$,
\begin{align*}
    \bar{M} &= \left[ \begin{array}{c} (I-N)(I+N)^{-1}\\ -2Z(I+N)^{-1} \end{array} \right], \\
    N &= X^TX + Z^TZ + Y - Y^T + \epsilon I.
\end{align*}

The approach outlined above, however, cannot be used directly to parameterize the state-space matrices of the perturbation model \eqref{eqn_CFpertsys}, which lacks a direct feedthrough term ($D = 0$). To circumvent this problem, in this work we employ a heuristic approach based on this parametrization that yields satisfactory results. Consider a system  ${G}_\theta$ with realization $({A}_\theta,{B}_\theta,{C}_\theta, D_\theta)$, where the state-space matrices are parameterized by $\theta=(d,\,V,\,X,\,Y,\,Z)$ using \eqref{eqn_ssparametrization}. During learning, we use the matrices $A_\theta$, $B_\theta$, and $C_\theta$ to define the perturbation model. Then, from the triangle inequality,~we~have
\begin{equation}
    \lVert [\Delta_N ~~ -\Delta_M] \rVert_\infty \leq \lVert {G}_\theta \rVert_\infty + \lVert {D}_\theta \rVert_2 \leq \gamma + \lVert {D}_\theta \rVert_2,
    \label{eqn_gamBnd}
\end{equation}
where $\lVert {D}_\theta \rVert_2 = \sqrt{\lambda_{max}({D}_\theta^T{D}_\theta)}$ is the operator norm of ${D}_\theta$. To solve the $\mathcal{H}_\infty$-norm-regularized learning problem, we update the loss function to
\begin{equation}
\mathcal{L}_\mathcal{D}(\eta,\,\theta,\,\alpha) = \mathcal{L}_{\text{pred}} + \beta_1 \mathcal{L}_{\text{dyn}} + \beta_2 \gamma + \beta_3 \lVert D_{\theta} \rVert_F,
\label{eqn_lossfn2}
\end{equation}
where the positive weights $\beta_1$, $\beta_2$, and $\beta_3$ are learning hyperparameters. The Frobenius norm of $D_{\theta}$ satisfies
\(\lVert D_{\theta} \rVert_F = \sqrt{\text{Tr}(D_{\theta}^TD_{\theta})} = \sqrt{\sum_{i} \lambda_i (D_{\theta}^T D_{\theta})} \geq \sqrt{\lambda_{max}(D_{\theta}^T D_{\theta})}\);
therefore, by minimizing $\gamma$ and $\lVert D_{\theta} \rVert_F$, we indirectly minimize the $\mathcal{H}_\infty$-norm of $ [\Delta_N ~~ -\Delta_M]$. When $\lVert D_{\theta} \rVert_F$ is non-zero, the upper bound on $\lVert  [\Delta_N ~~ -\Delta_M] \rVert_\infty $ provided in \eqref{eqn_gamBnd} is conservative. However, if $\lVert D_{\theta} \rVert_F$ is sufficiently small, we can infer that $\gamma$ serves as an approximate upper bound on $\lVert  [\Delta_N ~~ -\Delta_M] \rVert_\infty $ without being overly conservative. The upper bound $\gamma$ is parameterized as $e^\alpha$, with $\alpha \in \mathbb{R}$, to ensure it remains positive throughout the training process. 

\begin{table*}[t]
\renewcommand{\arraystretch}{1.3} 
\caption{Performance metrics of the learned norm-bounded coprime factor perturbations.}
\label{tab_learningResults}
\begin{center}
\begin{tabular}{c|c|c|c}
         & \textbf{Simple Pendulum} & \textbf{~~Van der Pol~~} & \textbf{~~3DOF UAS~~} \\ \hline
         $\mathcal{L}_{pred}$ (training) & $1.42\times 10^{-6}$ & $2.45\times 10^{-5}$ & $5.96\times 10^{-5}$ \\ \hline
         $\mathcal{L}_{pred}$ (validation) & $1.69\times 10^{-6}$ & $4.88\times 10^{-5}$ & $6.13\times 10^{-5}$ \\ \hline
         $\mathcal{L}_{dyn}$ (training) &  $2.27\times 10^{-6}$ &  $2.51\times 10^{-5}$ &  $2.45\times 10^{-5}$\\ \hline
         $\mathcal{L}_{dyn}$ (validation) & $2.54\times 10^{-6}$ & $3.70\times 10^{-5}$ & $2.50\times 10^{-5}$\\ \hline
         $\gamma$ & $0.565$ & $5.186$  & $0.817$ \\ \hline
         $\lVert D_\theta \rVert_F$ &  $8.46\times 10^{-5}$ & $1.20\times 10^{-3}$ & $4.93\times 10^{-4}$ \\ \hline
         $\lVert [\Delta_{N} ~~ -\Delta_{M}] \rVert_\infty$ & $0.558$ & $5.104$ & $0.745$\\      
    \end{tabular}
\end{center}
\end{table*}

\section{Numerical Examples}

We validate the proposed approach using three examples: a simple pendulum, a Van der Pol oscillator, and a 3-DOF longitudinal model of a UAS. The continuous-time differential equations governing the nonlinear dynamics of these systems are solved in MATLAB using \texttt{ode23} with some sampling time $\delta t$. The nominal system, $P$, is obtained by linearizing the nonlinear dynamics about the equilibrium of interest, and subsequently discretizing the resulting continuous-time LTI system using the zero-order hold method with a sampling time $\delta t$. The left coprime factor system $[-\tilde{N} ~~ \tilde{M}]$ corresponding to $P$ is defined as in \eqref{eqn_lcfP} with the stabilizing matrix $L_P$ generated using the MATLAB function $\texttt{dlqr}$. 
The learning problem is formulated and solved using the PyTorch-based DeepSI toolbox \cite{beintema2021nonlinear} with the Adam optimizer \cite{kingma2014adam}. All computations were performed on a desktop with 32 GB RAM and an Intel Xeon E-2224G 3.5 GHz CPU (4 cores). 

\vspace{2mm} \noindent \textbf{Simple pendulum:} The equations of motion for the simple pendulum are defined as
\begin{equation*}
    \dot{x}_1 = x_2, \quad \dot{x}_2 = 0.01x_2 - \sin{x_1} + u,
\end{equation*}
where $x = [x_1,\,x_2]^T$ is the state and $u$ denotes the input. State and input trajectories for learning are generated using open-loop nonlinear simulations, with the initial state $x_0$ randomly sampled from $[-\pi/3, ~ \pi/3] \times [-\pi/3, ~ \pi/3]$ and pseudorandom input sequences generated with a uniform distribution satisfying $\lvert u(k) \rvert \leq 0.5$. We conduct a total of $5000$ simulations, each spanning discrete-time instances $k=0,\,1,\,\dots,\,100$, with sampling time $\delta t=0.1~\mathrm{s}$. Of the $5000$ trajectories in the learning dataset, $80\,\%$ are used for training, and the remaining $20\,\%$ are used for validation. 

The dimension $N$ of the coprime factor perturbation system $[\Delta_{N} ~~ -\Delta_{M}]$ is set to $10$, and the weights in the loss function \eqref{eqn_lossfn2} are chosen as $\beta_1 = 0.1$, $\beta_2 = 10^{-5}$, and $\beta_3 = 10^{-5}$. The neural network parameterizing the lifting function $\Psi$ consists of two hidden layers, each containing $64$ neurons and followed by an ELU activation. The model is trained by minimizing the loss function over a prediction horizon length $T = 100$ across $5000$ epochs with a batch size of $256$.
Table~\ref{tab_learningResults} provides the training and validation losses of the learned system, along with the Frobenius norm of the feedthrough matrix $D_{\theta}$, which is obtained through the norm-bounded state-space parameterization \eqref{eqn_ssparametrization}, the value of the norm-bound $\gamma$, and the $\mathcal{H}_\infty$-norm of the learned system $\left[\Delta_N ~~ -\Delta_M\right]$.
As discussed in Section~\ref{sec_hinf}, minimizing $ \lVert D_{\theta} \rVert_F$ reduces the conservatism introduced by the triangle inequality \eqref{eqn_gamBnd}, thereby ensuring that $\gamma$ serves as a tight upper bound on $\lVert \left[\Delta_N ~~ -\Delta_M\right] \rVert_\infty$. 

\begin{figure*}[!t]
\centering
\subfloat[Simple Pendulum \label{fig_SpenState}]{\includegraphics[width=0.4\textwidth]{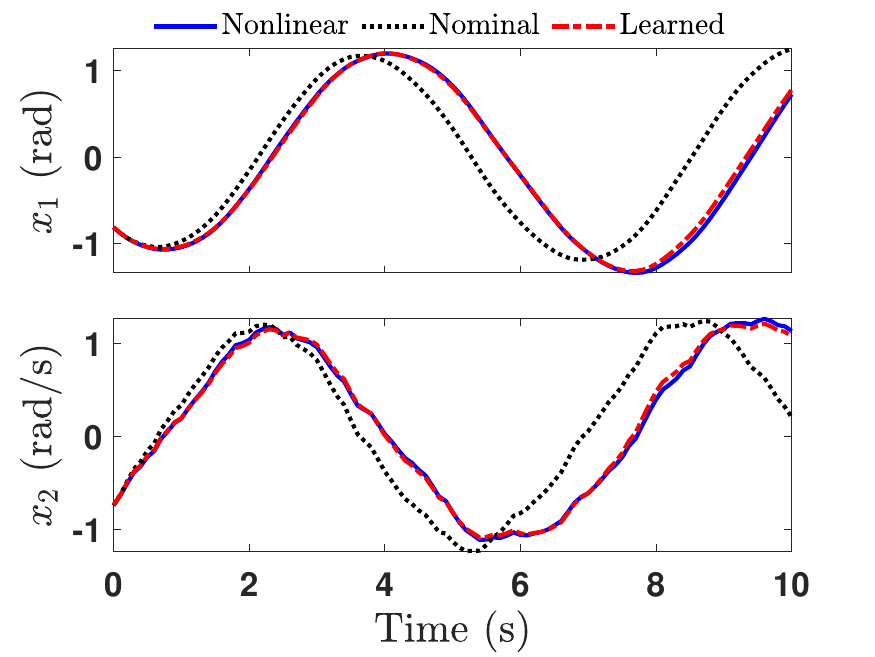}}\qquad
\subfloat[Van der Pol \label{fig_VpolState}] {\includegraphics[width=0.4\textwidth]{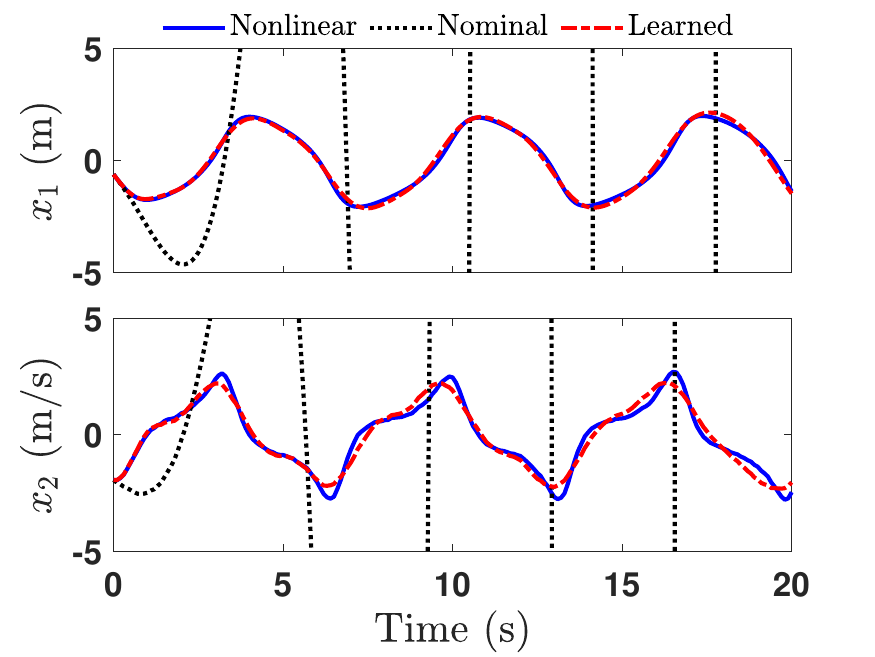}}
\caption{Output histories of the nonlinear and linear systems for the same input.}
\label{fig_State}
\end{figure*}

The linear approximation $G$ of the nonlinear dynamics is derived from the coprime factors $\tilde{N},\,\tilde{M}$ and the coprime factor perturbations $\Delta_N$,\, $\Delta_M$, as described in \eqref{eqn_sstransformation}. To demonstrate how discrepancy modeling leads to an improved linear approximation of the nonlinear system, we compare the output histories of the true nonlinear system, the nominal system $P$, and the learned LTI approximation $G$ for the same input history and initial state $x_0$. Recall that while the nonlinear and nominal systems share the same initial state, the initial state of the learned system $G$ is $[x_0^T,\,-\Psi(x_0)^T]^T$. Multiple test cases with randomly generated input histories and initial states within the previously defined envelopes are considered for comparison. The output histories of the three systems corresponding to a typical test case are shown in Figure~\ref{fig_SpenState}.

\vspace{2mm} \noindent \textbf{Van der Pol oscillator:} The equations of motion for the Van der Pol oscillator are given by
\begin{equation*}
    \dot{x}_1 = x_2, \quad \dot{x}_2 =  (1 - x_1^2)x_2 - x_1 + u.
\end{equation*}
State and input trajectories for learning are generated using open-loop nonlinear simulations, with the initial state $x_0$ randomly sampled from $[-2, ~ 2] \times [-2, ~ 2]$ and pseudorandom input sequences generated with a uniform distribution satisfying $\lvert u(k) \rvert \leq 0.5$. 
All other parameters, including the sampling time, number of simulations, lifted state space dimension, prediction horizon length, and learning hyperparameters, are identical to those used in the pendulum example. The learning results are summarized in Table~\ref{tab_learningResults}, while the output histories of the nonlinear, nominal, and learned systems from a typical simulation run are shown in Figure~\ref{fig_VpolState}. 

\begin{figure*}[tb]
\centering
\subfloat[Open-loop (same input) \label{fig_UASState}]{\includegraphics[width=0.4\textwidth]{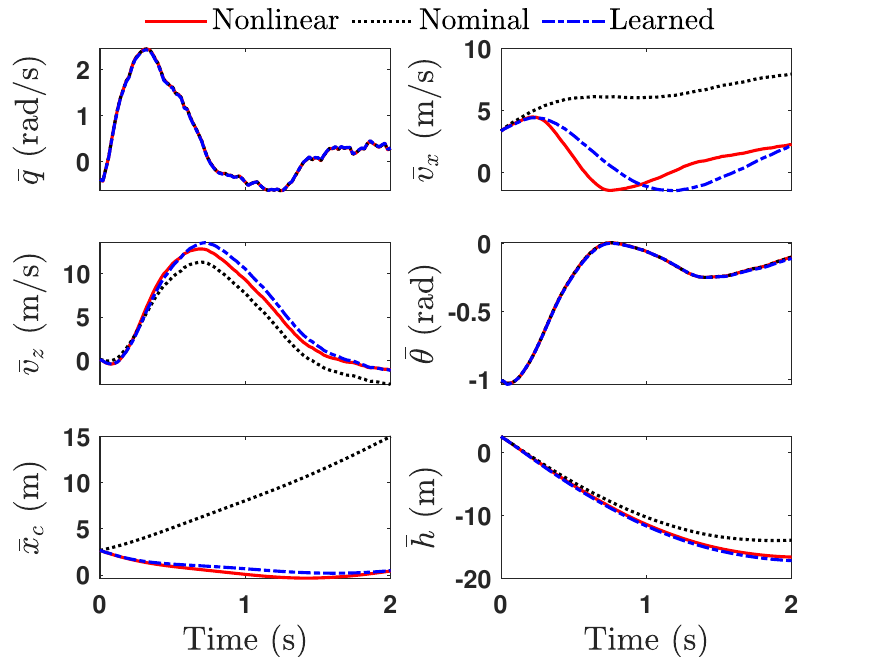}}\qquad
\subfloat[Closed-loop (same disturbances) \label{fig_UASStateCL}] {\includegraphics[width=0.4\textwidth]{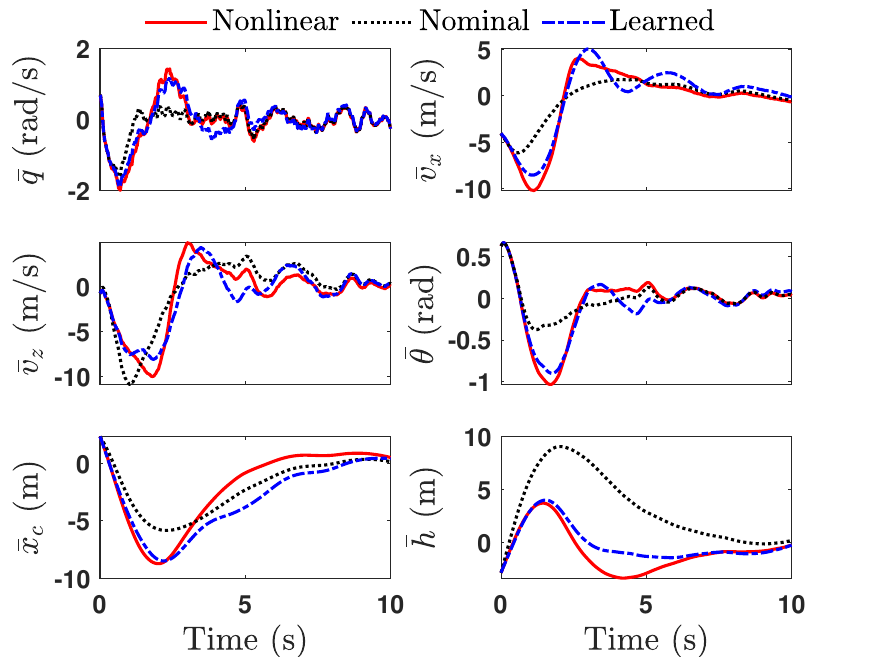}}
\caption{Output histories of the nonlinear and linear 3-DOF UAS models.}
\label{fig_UAS}
\end{figure*}

\vspace{2mm} \noindent \textbf{3-DOF UAS:} The dynamics of the 3-DOF longitudinal model of a UAS are described by 
\begin{equation*}
    \begin{bmatrix}
        \dot{q} \\ \dot{v}_x \\ \dot{v}_z \\ \dot{\theta} \\ \dot{x}_c \\ \dot{h}
    \end{bmatrix} =
    \begin{bmatrix}
        \tau/J \\ -qv_z + F_x/m - g\sin{\theta} \\ qv_x + F_z/m + g\cos{\theta} \\ q \\ v_x\cos{\theta} +  v_z\sin{\theta} \\ v_x\sin{\theta} -  v_z\cos{\theta}
    \end{bmatrix}
\end{equation*}
The equations can be written in state-space form with $x = [q,\,v_x,\,v_z,\,\theta,\,x_c,\,h]^T$ as the state and $u = [F_x,\,F_z,\,\tau]^T$ as the input. The state variables $q,\,v_x,\,v_z,\,\theta,\,x_c,\,h$ correspond to the aircraft's pitch rate, velocity along body x-axis, velocity along body z-axis, pitch angle, forward position in the inertial frame, and altitude, respectively. The input variable $\tau$ denotes the applied pitching moment, while $F_x,\,F_z$ are the forces applied in the forward and downward directions, respectively. The constant $m = 5.71~\mathrm{kg}$ is the mass of the UAS, and $J=1.57~\mathrm{kg\text{-}m^2}$ is its moment of inertia.

We are interested in approximating the behavior of the UAS around a non-zero trim $(x^\ast(k),\,u^\ast)$, corresponding to level flight at an airspeed of $15~\mathrm{m/s}$. The trim state is time-dependent as the reference position changes over time.  
The nominal linearized model remains LTI because the UAS dynamics are independent of $x_c$. For this example, we conduct a total of $10000$ nonlinear simulations, each spanning discrete-time instances $k=0,\,1,\,\dots,\,100$, with sampling time $\delta t=0.02~\mathrm{s}$.
The UAS can quickly exhibit unstable behavior under random inputs. Therefore, to generate the learning dataset, we conduct closed-loop simulations using a full-state feedback controller designed to regulate the UAS about the trim. 
During simulations, the initial state $x_0$ is sampled from the following hyperrectangle: 
$\{ p = (p_1,\,\dots,\,p_6) \in \mathbb{R}^6 ~|~ \lvert p_i \rvert \leq \pi/3, \lvert p_j \rvert \leq 5, ~\text{for}~ i = 1,4 ~\text{and}~j=2,3,5,6 \}$. 
To generate a rich dataset, we introduce exogenous disturbances in the form of measurement noise, which corrupts the state measurements, and process noise, which affects the input commands generated by the controller. The measurement noise is sampled from a normal distribution with a zero mean and a standard deviation of $0.15$ for pitch angle and rate measurements, and $2$ for position and velocity measurements. The process noise is sampled from a uniform distribution with upper bounds of $5~\mathrm{N}$, $10~\mathrm{N}$, and $2~\mathrm{N\text{-}m}$ for the absolute values of  $F_x$, $F_z$, and $\tau$, respectively.

Before learning, we apply a change of variables: $\bar{x} = x - x^\ast$ and $\bar{u} = u - u^\ast$ to the dataset to ensure that the learned coprime factor perturbations have an equilibrium at the origin. The dimension $N$ of the coprime factor perturbation system is set to $20$, and the weights in the loss function \eqref{eqn_lossfn2} are chosen as $\beta_1 = 0.1$, $\beta_2 = 10^{-5}$, and $\beta_3 = 10^{-2}$. All other parameters remain the same as those used in the previous examples. The learning results are summarized in Table~\ref{tab_learningResults}, and the output histories of the nonlinear, nominal, and learned systems from a representative open-loop simulation run are shown in Figure~\ref{fig_UASState}. For the same input, the pitch angle and pitch rate outputs are nearly identical across all three cases, as their governing equations are linear.
Since the system is unstable, the remaining outputs of the linear approximations eventually diverge from those of the nonlinear system over a longer horizon due to the exponential amplification of errors over time. Nevertheless, system $G$ demonstrates a significantly delayed divergence compared to $P$, owing to its improved approximation accuracy. Furthermore, we compare the outputs of the three systems in a closed-loop simulation setup where the controller, initial conditions, and disturbances are identical across all systems. The output histories from a representative closed-loop simulation run are shown in Figure~\ref{fig_UASStateCL}. Notably, the learned system's response aligns more closely with the nonlinear system's behavior compared to the nominal model. 

\begin{figure}[t]
\begin{center}
\centerline{\includegraphics[width=0.49\textwidth]{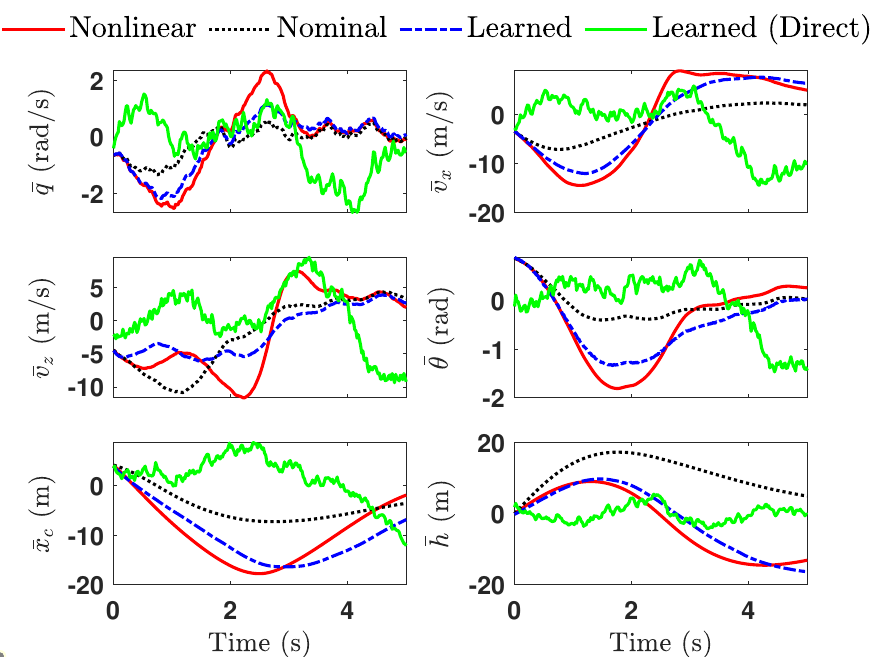}}
\caption{Comparison of output histories of the nonlinear UAS model with those of the nominal linear model and the linear lifted  models in a closed-loop setting.}
\label{fig_UASStateCL_2}
\end{center}
\end{figure}

As discussed in Section~\ref{sec_intro}, a purely data-driven approach can be used to directly learn the linear lifted approximation $G$. Specifically, the matrices $A_G$, $B_G$, $C_G$, and the lifting function $\Phi$ can be learned using an RNN architecture similar to the one shown in Figure~\ref{fig_RNN}. However, the discrepancy modeling approach, which leverages the nominal model, proves advantageous especially when the available training data is limited. To demonstrate this, we generate $64$ nonlinear system trajectories, each $5~\mathrm{s}$ in duration. These trajectories correspond to all possible combinations of the six initial state variables, with each variable set to one of its extreme values. Linear lifted models are then learned from this dataset using both the discrepancy modeling and direct learning approaches. Figure~\ref{fig_UASStateCL_2} shows the closed-loop responses of the nonlinear system, the nominal linear model, and the lifted models. As observed from the plots, the directly learned model performs poorly in approximating the nonlinear system dynamics, which is expected given the sensitivity of learning-based approaches to training data. In contrast, the lifted model obtained via the discrepancy modeling approach offers a significantly better approximation, also surpassing the nominal model. This improvement stems from the use of the nominal model as a baseline, which is then enhanced by the data-driven approach.

\section{Conclusions and Future Work}

This work presents a coprime factorization-based approach to model the discrepancies between a nonlinear system and its nominal linear approximation using norm-bounded dynamic perturbations. A deep learning method is employed to learn these dynamic perturbations in a higher-dimensional state space, while a direct parametrization is utilized to minimize the norm of the learned perturbations. Through multiple examples, we demonstrate the effectiveness of the proposed approach in generating linear models that more closely approximate the nonlinear system's behavior, offering significant improvements over traditional linearization-based approximations.

While this work focuses on modeling, control design for nonlinear systems based on the learned models will be addressed in future work.
Coprime factorization is also linked to important concepts in robust stabilization, such as stability margins and gap metrics. The Hankel-norm of the normalized left coprime factors of an LTI system provides the optimal stability margin, which measures the degree of uncertainty the system can tolerate before becoming unstable in a closed-loop configuration. Future work will explore the integration of these concepts into the proposed approach.

\addtolength{\textheight}{-12.2cm}  
\bibliography{ieeeconf}
\bibliographystyle{IEEEtran}

\appendix
\section{Direct Approximation of the Coprime Factors of $G$} \label{sec_app}

Consider the scenario where the nominal LTI approximation of the nonlinear dynamics is unavailable. Existing approaches \cite{korda2018linear,proctor2018generalizing,narasingam2019koopman,folkestad2020extended,mamakoukas2019local} can be utilized to directly learn the LTI approximation $G$, as defined in \eqref{eqn_LTIsys}, of the nonlinear system. However, we propose an alternative coprime factorization-based method, where instead of directly learning the system $G$, we focus on learning its left coprime factorization as defined by \eqref{eqn_lcfss}.
Unlike the training of unstable systems, where a short prediction horizon may be necessary, the coprime factors, which are inherently stable, can be trained over longer horizons.

The coprime factorization can be viewed as the sum of an unknown dynamical system $G_\theta$ and a known static system with feedthrough matrix $G_I = [0 ~~ I]$. The dynamical system is described as
\begin{equation*}
 {G}_\theta  \colon \begin{cases}
     \tilde{z}(k+1) = A_\theta \tilde{z}(k) + B_\theta \tilde{u}(k),  ~\, \tilde{z}(0) = \Psi(x_0),\\
     \hspace{3.2mm} -\hat{x}(k) = C_\theta \tilde{z}(k),
  \end{cases}
\end{equation*}
where $\Psi : \mathbb{R}^n \rightarrow \mathbb{R}^N$ is the lifting function and $\hat{x}$ is an approximation of the true nonlinear system state $x$. 
This system can be learned by minimizing a loss function of the form \eqref{eqn_lossfn}, where 
$\mathcal{L}_{\text{pred}}$ quantifies the MSE between $x$ and $\hat{x}$, and $\mathcal{L}_{\text{dyn}}$ quantifies the MSE between $\tilde{z}$ and $\Psi(x)$.
The stability of the system can be enforced by utilizing the direct parametrization proposed for state matrices. The state-space matrices of $G$ are recovered as $(A_G,\,B_G,\,C_G) = (A_\theta - B_{\theta_2}C_\theta,\, -B_{\theta_1}, \,C_\theta)$, where $B_{\theta_1}$ and $B_{\theta_2}$ are obtained by partitioning $B_{\theta}$. Additionally, we have $z(0) = -\Psi(x_0)$.

\end{document}